\documentstyle{article}
\newcommand{\be}{\begin{equation}}
\newcommand{\bea}{\begin{eqnarray}}
\newcommand{\ba}{\begin{array}}
\newcommand{\bean}{\begin{eqnarray*}}
\newcommand{\ee}{\end{equation}}
\newcommand{\eea}{\end{eqnarray}}
\newcommand{\ea}{\end{array}}
\newcommand{\eean}{\end{eqnarray*}}
\newcommand{\hPi}{\mbox{$\widehat{\Pi}$}}
\newcommand{\hGa}{\mbox{$\widehat{\Gamma}$}}

\newcommand{\hA}{\mbox{$\hat{A}$}}

\newcommand{\hD}{\mbox{$\hat{D}$}}

\newcommand{\hSi}{\mbox{$\hat{\Sigma}$}}
\newcommand{\hm}{\mbox{$\hat{m}$}}
\newcommand{\bs}{\mbox{$\bar{s}$}}
\newcommand{\hs}{\mbox{$\hat{s}$}}

\newcommand{\WIs}{Ward identities~}

\def \dsl {\partial \kern-.55em{/}}
\def \Dsl {D \kern-.65em{/}}
\def \qsl {q \kern-.45em{/}}
\def \slp {p \kern-.45em{/}}
\def \ksl {k \kern-.45em{/}}

\def \S {S-matrix}
\def \gV {\mbox{$\gamma_5$}}
\def \cons {contributions~}

\def \se {self-energy~}
\def \ses {self-energies~}
\def \sesc {self-energies,}
\def \sesp  {self-energies.}
\def \ind {independent~}
\def \ol {one-loop~}
\def \Den {\mbox{{\large D}}}

\newcounter{subeq}
\begin{document}
\begin{titlepage}
\rightline{ NYU-TH-95/10/01}
\rightline{October 1995}

\vspace*{1cm}
\Large
\begin{center}
{\bf  Application of the Pinch Technique to Neutral Current
Amplitudes and the Concept of the $Z$ Mass }
\end{center}

\vspace*{1cm}
\large
\begin{center}
{\bf Kostas Philippides and Alberto Sirlin }
\end{center}

\vspace*{1cm}
\large
\begin{center}
           {\it New York University, Department of Physics

      Andre and Bella Meyer Hall of Physics, 4 Washington Place

                    New York,NY 10003,USA.}
\end{center}

\vspace*{1cm}
\begin{center}
\large
{\bf ABSTRACT}
\end{center}
\vspace*{0.5cm}
\baselineskip=23pt

The pinch technique (PT) is applied to neutral current amplitudes,
focusing on the mixing problem. Extending recent arguments due to
Papavassiliou and Pilaftsis, it is shown that the use of the PT
self-energies does not shift the complex-valued position of the
pole through order {\cal O}($g^4$). This leads (to the same accuracy)
to a simple interpretation of $M_Z$, the mass measured at LEP, in
terms of the PT self-energies. It is pointed out that the PT approach
provides a convenient and rather elegant formalism to discuss important
neutral current amplitudes, such as those relevant to four-fermion
processes and LEP2.

\end{titlepage}

\large
\baselineskip=20pt

\setcounter{equation}{0}
\section{Introduction}

The Pinch Technique (PT) is an algorithm that automatically re-arranges
\S~ \cons into \ses, vertex and box diagrams which are separately
$\xi$-independent and satisfy very desirable theoretical properties
[1,2].
 We emphasize that this re-arrangement occurs naturally {\it before}
any calculation is performed (one-loop integrations or Dirac algebra)
by virtue of the naive Ward identities satisfied by the
tree level vertices of the theory.
This fact is particularly  transparent in the formulation of the PT
in terms equal time commutators of currents \cite{DeSi}.
A temporary drawback has been
that so far the PT has been fully developed only at the one-loop level.
Very recently, however, Papavassiliou and Pilaftsis (P-P) have been able to
extend to higher orders certain aspects of the construction of the PT \ses
in charged current processes \cite{PaPi}.
In particular, they have obtained the
important result that the use of the PT \ses does not shift the
complex-valued position of the pole, as determined from the conventional
\ses$\!$. This opens the door for a number of significant developments. For
instance, P-P have proposed to use their PT formalism to discuss resonant
amplitudes involving unstable particles.

In section 2 of this paper we apply the PT to the phenomenologically
important neutral current amplitudes, focusing on the mixing problem.
In section 3 we extend the P-P argument to the neutral current case and show
that the position of the complex-valued pole is not shifted through
\cal{O}($g^4$). This permits us to obtain a very simple theoretical
interpretation of the $Z$ mass, measured at LEP. Specifically, through
\cal{O}($g^4$),
\be
M_Z^2 = M_{0}^2 + \Re e\hA(M_Z^2) ~,
\label{zmass}
\ee
where  $\hA$ is the PT transverse self-energy of $Z$, including the
contributions of tadpoles and $\gamma Z$ mixing
effects that start in \cal{O}($g^4$), and $M_0$ is the gauge invariant
combination of the bare $Z$ mass and the tadpole counterterms.
In Section 3 we also remind the
reader of the theoretical difficulties that arise if, as it was done for
a long time, one attempts to define the renormalized mass in terms of
Eq.(\ref{zmass}) with $\hA$ replaced by the conventional \se $A$.
Eq.(\ref{zmass}) tells us that, through \cal{O}($g^4$), $M_Z$ can be
identified with the zero of the real part of the inverse transverse
propagator, provided that this is constructed with the PT \ses$\!$.
In Section 3 we also point out that the PT approach provides a convenient
and rather elegant formalism to discuss neutral current amplitudes such
as those  relevant to four-fermion processes and LEP2.
An important advantage of this formalism is that
it treats bosonic and fermionic \cons on an equal footing.

\setcounter{equation}{0}
\section{Application of the PT to neutral current amplitudes}
In this section we discuss the application of the PT to the
phenomenologically important neutral-current amplitudes, focusing
on the mixing problem and restricting the discussion to one-loop order.

Fig. 1 depicts  the 16 \ol \se graphs in the Standard Model (SM) contributing
to neutral current processes such as $e^+e^- \rightarrow f\bar{f}$, where
$f$ is a generic fermion distinct from $e$, and
 $e^+e^- \rightarrow W^+W^-$, which will be important for LEP2 studies.
In the figures $\chi$ is the unphysical Goldstone boson associated with
$Z$ and it is understood that the blobs and lines stand for the conventional
\ses and tree level propagators in the $R_{\xi}$ gauges. Thus, for example,
the photon propagator, $i\Delta^{\mu\nu}_{\gamma}$, is given by
\be
i\Delta^{\mu\nu}_{\gamma} = -i
\left[ t^{\mu\nu}+\xi_{\gamma}\ell^{\mu\nu}\right]/q^2 ~,
\label{gprop}
\ee
where $t^{\mu\nu} \equiv g^{\mu\nu}-q^{\mu}q^{\nu}/q^2$ and
$\ell^{\mu\nu} \equiv q^{\mu}q^{\nu}/q^2$,
and the $Z$ propagator, $i\Delta^{\mu\nu}_{Z}$, by
\be
i\Delta^{\mu\nu}_{Z} = -i\left[g^{\mu\nu}
-(1-\xi_Z)q^{\mu}q^{\nu}\Delta_{\chi}\right]/(q^2-M_0^2) ~,
\label{Zprop}
\ee
where
$\Delta_{\chi}\equiv (q^2-\xi_ZM_0^2)^{-1}$.
We also have the equivalent expression
\be
\Delta^{\mu\nu}_{Z} = U^{\mu\nu}_{Z} -(q^{\mu}q^{\nu}/M_0^2)
\Delta_{\chi}~,
\label{Zpropequiv}
\ee
where
\be
U^{\mu\nu}_{Z}
=
-t^{\mu\nu}/(q^2-M_0^2)+\ell^{\mu\nu}/M_0^2
\label{unitprop}
\ee
corresponds to the propagator in the unitary gauge.

In the application of the PT the first step is to replace the conventional
\ses in Fig.1 by their PT counterparts. This is done by combining the
conventional amplitudes with the one-loop pinch parts from vertex and box
diagrams. The result are new, $\xi$-\ind \ses$\!$, that satisfy very desirable
theoretical properties. As the answer is $\xi$-\ind$\!$, a convenient
short-cut is to work from the outset in the $\xi_i=1$ gauge. In that case
there are no pinch parts emerging from the propagators and, therefore,
when one considers four-fermion
processes, there are no pinch parts at one loop
arising from box diagrams.
Once the \ses in Fig.1 have been replaced by their PT  counterparts, there
remains the $\xi$ dependence of the tree level propagators in the $R_{\xi}$.
This dependence, however, can be shown to cancel
by employing the elementary Ward identities
satisfied by the PT \ses \cite{WI}. Defining the \ses of vector
bosons, scalar bosons, and mixed vector-scalar bosons, as the corresponding
one-particle irreducible Feynman diagrams multiplied by $-i,~i,$ and 1,
respectively, and denoting the PT \ses by caret amplitudes, we have \cite{WI}
\be
 q^{\mu}\hPi^{\gamma\gamma}_{\mu\nu} =
q^{\mu}\hPi^{\gamma Z}_{\mu\nu} =
q^{\mu}\hPi^{\gamma \chi}_{\mu} =
q^{\mu}\hPi^{\gamma H}_{\mu} = 0 ~,
\label{WI1}
\ee
\be
q^{\mu}\hPi^{ZZ}_{\mu\nu} + M_0 \hPi_{\nu}^{\chi Z} = 0  ~,
\label{WI2}
\ee
\be
q^{\mu}\hPi^{ZH}_{\mu} + M_0 \hPi^{\chi H} =  0 ~,
\label{WI3}
\ee
\be
q^{\mu}q^{\nu}\hPi^{ZZ}_{\mu\nu} + M_0^2 \hPi^{\chi\chi} = 0 ~.
\label{WI4}
\ee

Employing these \WIs it is straightforward  to show that the remaining
$\xi$-dependence in the propagators cancels and the \se graphs
can be split into two groups corresponding to transverse and longitudinal
\sesc each group containing four amplitudes.
Decomposing, for instance,
\be
\hPi_{\mu\nu}^{ZZ} = t_{\mu\nu}\hPi_T^{ZZ} + \ell_{\mu\nu}\hPi_L^{ZZ}~,
\label{ZZdec}
\ee
where the subindices $T$  and $L$ mean ``transverse'' and ``longitudinal'',
the transverse and longitudinal amplitudes are depicted in Fig.2(a) and
Fig.2(b),
respectively.
The propagators in Fig.2(a) are $-i/(q^2-M_0^2)$ and $-i/q^2$ ;
in Fig.2(b) they are $i/M_0^2$ and $-i/(q^2-m_0^2)$ where
$m_0$ is the bare Higgs mass.
We note  that the decomposition involves only the
physical particles,
$\gamma,~Z,~H,$ with $\xi$-\ind propagators and self-energies.
It is worth noting that the
only surviving $q^2=0$ poles are associated with the photon exchange
diagram of Fig.2(a). For example, $\hPi^{\gamma Z} \propto q^2$ and
$q_{\nu}J^{\nu}_{\gamma}=0$ which cancel the $1/q^2$ singularity in
the $\gamma Z$ mixing diagrams in Fig.2(a).
As $q^2 \rightarrow 0$, the sum of the $ZZ$ diagrams in
Fig.2(a) and Fig.2(b) includes
$(q_{\mu}q_{\nu}/q^2) [\hPi_{L}^{ZZ}(q^2)-\hPi_{T}^{ZZ}(q^2)]$. Recalling that
$\hPi_{\mu\nu}^{ZZ}$ is regular as  $q^2 \rightarrow 0$, we see from Eq.(10)
that in this limit $ \hPi_{L}^{ZZ}(q^2)-\hPi_{T}^{ZZ}(q^2) \propto q^2$,
which cancels the $1/q^2$ singularity.

P-P have proposed a method to construct iterated chains of PT
\ses \cite{PaPi}.
This will be illustrated in
the $Z$ case in Section 3. For the moment we point out that the iteration
of the transverse PT \ses can be summed up by the same procedure employed
for ordinary \sesp~ Specifically, one inverts the matrix \cite{BaCo}
\be
M =
\left( \begin{array}{cc}
               q^2 - M_0^2 - \hPi_T^{ZZ} &- \hPi^{\gamma Z}\\
                       - \hPi^{\gamma Z} &q^2 - \hPi^{\gamma\gamma}
       \end{array}
\right) ~.
\label{matrixgZ}
\ee
The inverse matrix is
\be
M^{-1} = \frac{1}{D (q^2-\hPi^{\gamma\gamma})} \left( \begin{array}{cc}
                  q^2-\hPi^{\gamma\gamma} & \hPi^{\gamma Z} \\
\hPi^{\gamma Z}&
q^2 - M_0^2 - \hPi_T^{ZZ}
 \end{array}
\right) ~,
\label{invmatrixgZ}
\ee
where $D\equiv  q^2 - M_0^2 - \hPi_T^{ZZ} -
(\hPi^{\gamma Z})^2/(q^2-\hPi^{\gamma\gamma})$.
The physical meaning of Eq.(\ref{invmatrixgZ})
becomes more transparent if we consider the
matrix element
$(\hGa_Z , \hGa_{\gamma})_{\mu}t^{\mu\nu}  M^{-1}
(\hGa'_Z , \hGa'_{\gamma})^T_{\nu} ,$
where
$\hGa_Z^{\mu},~$$\hGa'^{\mu}_Z ,~$$\hGa^{\mu}_{\gamma}
,~$$\hGa'^{\mu}_{\gamma}$ represent
appropriate vertex functions of $Z$ and $\gamma$ with external particles.
One readily finds \cite{Stu}
\be
(\hGa_Z , \hGa_{\gamma})^{\mu}t_{\mu\nu} M^{-1}
\left( \begin{array}{c} \hGa'_Z \\ \hGa'_{\gamma}
\end{array}\right)^{\nu} =
[\hGa_{Z} + \hGa_{\gamma}
\frac{\hPi^{\gamma Z}} {q^2-\hPi^{\gamma\gamma}}]^{\mu}~
\frac{t_{\mu\nu}}{\Den}~
[\hGa'_{Z} + \hGa'_{\gamma}
\frac{ \hPi^{\gamma Z} } { q^2-\hPi^{\gamma\gamma}}]^{\nu}
 + \hGa^{\mu}_{\gamma}
\frac{ t_{\mu\nu} }{ q^2-\hPi^{\gamma\gamma} }
\hGa'^{\nu}_{\gamma} ~.
\label{Tr}
\ee
Eq.(\ref{Tr}) can be interpreted as describing the propagation of a $Z$ with
dressed propagator $-it_{\mu\nu}/D$ between the
sources $\hGa^{\mu}_{Z}$ and $\hGa'^{\nu}_{Z}$ and
of a $\gamma$ with dressed propagator $-it_{\mu\nu}/(q^2-\hPi^{\gamma\gamma})$
between $\hGa^{\mu}_{\gamma}$ and $\hGa'^{\nu}_{\gamma}$.
It is interesting to note that this last term
contains the effect of the running of the electromagnetic coupling
constant $\alpha$ involving both its fermionic and bosonic contributions
\cite{DeSi}.
The remaining terms in Eq.(\ref{Tr}) represent $\gamma -Z$
 mixing effects.

Considering now the longitudinal part of a four-fermion amplitude,  we use
$
\hPi_L^{ZZ}=-(M_0^2/q^2)\hPi^{\chi\chi},~
\hPi^{ZH} = -(M_0/q^2)\hPi^{\chi H}
$
where $\hPi^{ZH}$ is defined by $\hPi^{ZH}_{\mu}=q_{\mu}\hPi^{ZH}$,
and we have employed  Eq.(\ref{WI3}) and  Eq.(\ref{WI4}).
Using also  $\partial_{\mu}J_Z^{\mu}  =  -M_0 J_{\chi}$,
where $J_Z^{\mu}$ and $J_{\chi}$ are the fermionic operators coupled
to $Z$ and $\chi$,  the \se matrix to be
inverted is given by
\be
L = \left( \begin{array}{cc}
                        q^2-\hPi_{\chi\chi}&  \hPi_{\chi H}\\
                          \hPi_{\chi H} &q^2 - m_0^2 - \hPi_{HH}
       \end{array}
\right)
\Rightarrow
L^{-1} =  \left( \begin{array}{cc}
                        \hD_{\chi\chi}&\hD_{\chi H}\\
                        \hD_{\chi H} &\hD_{HH}
       \end{array}
\right)~,
\ee
where $m_0$ is the Higgs bare mass and,
$
\hD_{\chi\chi} =[q^2-\hPi_{\chi\chi}-\frac{\hPi_{\chi H}^2}
{q^2 - m_0^2 - \hPi_{HH}}]^{-1}  $,
$\hD_{HH} =[q^2-m_0^2-\hPi_{HH}-\frac{\hPi_{\chi H}^2}
                                  {q^2-\hPi_{\chi\chi}}]^{-1}
$, $
\hD_{\chi H} = -\hPi_{\chi H}[
     (q^2-\hPi_{\chi\chi})
       (q^2-m_0^2-\hPi_{HH})-(\hPi_{\chi H})^2]^{-1}~.$
Thus, in this part of the amplitude all longitudinal $Z$'s can be replaced
by $\chi$'s, their unphysical scalar counterparts,
for arbitrary values of $q^2$.

We next consider  the one-loop vertex diagrams which  provide
pinch parts to the \ses and also receive vertex-like pinch parts
from the boxes. They are trasformed into  new
$\xi$-\ind expressions which satisfy a number of desirable
properties. i) they are UV finite,  ii) the form factors extracted from
them and associated
with the various electromagnetic and weak moments of the external particles
are IR finite and well behaved for $q^2 \rightarrow \infty $
(they respect perturbative unitarity) \cite{FF}
 and iii) the new one loop vertices
satisfy their tree level Ward identities. Because of the WI
\be
q^{\mu}\hGa^{Z\bar{f}f}_{\mu} + iM_0\hGa^{\chi\bar{f}f} =
(g/2c_W)\left[(g_V^f+g_A^f\gV)\hSi_f(p)-\hSi_f(p+q)(g_V^f-g_A^f\gV)
\right]~,
\label{VWI}
\ee
the graphs with the $\chi\bar{f}f$ vertices cancel
against longitudinal $\xi-$dependent contributions from $Z$ exchange
(cf. last term in Eq.(\ref{Zpropequiv})). In this way one is left with graphs
that contain only physical particles.
Finally we turn our attention to the one-loop box graphs.
In the case of  four fermion processes,
the remainder of the box diagrams after the PT subtraction
is their expression in the
t' Hooft-Feynman gauge ($\xi_i=1$), which is UV finite.
For arbitrary $\xi_i$,   when the external
fermions are considered massless,  the $ZZ$, $Z\gamma$ and $\gamma\gamma$
boxes are gauge independent while the $WW$ box  gives pinch parts only
to the self-energies.
If the  external fermions  have a finite mass,
then the box graphs also contribute pinch terms to the vertices.
For $e^+e^-\rightarrow W^+W^-$,  the PT expression for the
boxes does not coincide with the one in the $\xi_i=1$  gauge, but it is still
UV convergent.

\setcounter{equation}{0}
\section{ Residual terms in $\hPi^{ZZ}_T$ and the definition of the $Z$ mass.}
As mentioned in Section 2, P-P have proposed the construction of chains
of  PT   \sesp~  To two and higher loops, they have shown that this
requires not only the pinch parts from the available vertex graphs
at the ends of the chain,  but also additional pinch
contributions from the one-particle
irreducible \ses of higher order than the ones contained in the chain.
An important result of their analysis is the demonstration
that the use of the PT \ses does not displace the complex-valued position
of the pole. In this section we extend the P-P argument to neutral-current
amplitudes   taking into account the effect of mixing. We restrict ourselves
to terms up to \cal{O}($g^4$), as this is sufficient for our purposes.
After proving that the position of the complex pole is not displaced through
\cal{O}($g^4$), we show how this leads to Eq.(\ref{zmass}), which provides a
simple field-theoretical definition of the $Z$ mass measured at LEP, and
permits to carry out the conventional mass renormalization.

We begin by considering the two-loop diagrams depicted in Fig.3. Recalling
\cite{DeSi}
\bea
\hPi^{ZZ}_T = &  \Pi^{ZZ}_T|_{\xi_i=1} - (q^2-M_0^2)4g^2c_W^2I_{WW}(q^2)~,
\label{PTZZ}\\
\hPi^{\gamma Z} = &  \Pi^{\gamma Z}_T|_{\xi_i=1} - (2q^2-M_0^2)2g^2c_Ws_W
I_{WW}(q^2)~,
\label{PTgZ}\\
\hPi^{\gamma\gamma} = &  \Pi^{\gamma\gamma}_T|_{\xi_i=1}
- (q^2-M_0^2)4g^2s_W^2I_{WW}(q^2)~,
\label{PTgg}
\eea
where $s^2_W=1-c_W^2$ is an abbreviation for $\sin\theta_W^2$, and
\be
I_{WW}(q^2) = i\mu^{4-n}\int \frac{d^nk}{(2\pi)^n}
\frac{1}{\left(k^2-M_W^2\right)\left[(k+q)^2-M_W^2\right]} ~,
\label{Iww}
\ee
the diagrams in Fig.3(a) are equal to
\bea
&i(g/c_W)^2
J_Z^{\mu}t_{\mu\nu}J_Z^{\nu}/(q^2-M_0^2)^2
\left\{
\left[\Pi^{ZZ}_T|_{\xi_i=1} - (q^2-M_0^2)4g^2c_W^2I_{WW}(q^2)\right]^2
/(q^2-M_0^2)
\right. \nonumber \\
 &~~~~~~~~~~~~~~\left. +
\left[ \Pi^{\gamma Z}_T|_{\xi_i=1}
- (2q^2-M_0^2)2g^2s_Wc_WI_{WW}(q^2) \right]^2/q^2 \right\}~.
\label{2PT}
\eea
On the other hand, the ordinary \ses plus the pinch parts of Fig.3(b,c,d)
proportional to $J_Z^{\mu}t_{\mu\nu}J_Z^{\nu}$, both  evaluated in the
$\xi_i=1$ gauge, give only
\bea
&i\left(\frac{g}{c_W}\right)^2 J_Z^{\mu} \frac{t_{\mu\nu}}{(q^2-M_0^2)^2}
 J_Z^{\nu}
\left\{ \frac{1}{q^2-M_0^2}\left[\Pi^{ZZ}_T|_{\xi_i=1}\right]^2
+\frac{1}{q^2}\left[ \Pi^{\gamma Z}_T|_{\xi_i=1} \right]^2
-4g^2c_W^2\Pi^{ZZ}_T|_{\xi_i=1}I_{WW} \right.~~~\label{2avail} \\
& \left. -\frac{q^2-M_0^2}{q^2}4g^2c_Ws_W\Pi^{\gamma Z}_T|_{\xi_i=1} I_{WW}
+(q^2-M_0^2)4g^4c_W^4I_{WW}^2 + \frac{(q^2-M_0^2)^2}{q^2}4g^4c_W^2s^2_WI_{WW}^2
\right\} ~. \nonumber
\eea
The additional  pinch terms that should be added to
the chain of Eq.(\ref{2avail})
in order to convert it into the  PT chain of
Eq.(\ref{2PT}) are readily obtained from their
difference  and are given by
\bea
R^{(2)}_{ZZ} =& -4g^2\left( c_W^2\Pi^{ZZ}_T|_{\xi_i=1} +
c_Ws_W \Pi^{\gamma Z}_T|_{\xi_i=1}\right) I_{WW}
\hspace*{2cm}
 \nonumber \\
&+(q^2-M_0^2)12g^4c_W^4I_{WW}^2 + (3q^2-2M_0^2)4g^4s^2_Wc^2_WI_{WW}^2 ~,
\label{R2ZZ}
\eea
where the common factor containing the currents and the two propagators
has been ommited.
We observe that  $R^{(2)}_{ZZ}$, in contrast to
the amplitudes  between curly brackets in
Eq.(\ref{2PT}) and Eq.(\ref{2avail}),
contains no propagators and thus
is of the same form as the
two-loop one-particle irreducible graphs of the $Z$  \se $\hPi^{(2)ZZ}_T$.
Following the P-P approach, one adds $R^{(2)}_{ZZ}$ to Eq.(\ref{2avail})
and subsequently subtracts  the same amplitude from  $\hPi^{(2)ZZ}_T$.

For completeness,  we also give the residual  two-loop terms for the
photon and the mixed \ses :
\bea
R^{(2)}_{\gamma\gamma } =& -4g^2s_W^2\Pi^{\gamma \gamma}|_{\xi_i=1} I_{WW}
 -4g^2c_Ws_W\Pi^{\gamma Z}_T|_{\xi_i=1}I_{WW} \hspace*{1.5cm} \nonumber \\
& + 12g^4s_W^4q^2 I_{WW}
 + 4g^4c_W^2s_W^2(3q^2-M^2_Z)I^2_{WW} \hspace*{0.5cm} ~,
\label{R2gg}
\eea
\bea
R^{(2)}_{\gamma Z } = &
-2g^2c_Ws_W\left(\Pi^{\gamma \gamma}|_{\xi_i=1} + \Pi^{ZZ}_T|_{\xi_i=1}
  \right)I_{WW} \hspace*{3cm}\nonumber \\
& -2g^2 \Pi^{\gamma Z}_T|_{\xi_i=1} I_{WW}  +
4g^4c_Ws_W[3q^2-(1+c_W^2)M^2_Z]I_{WW}^2
\label{R2gZ}~.
\eea

Since the mid-eighties a number of authors have proposed the idea that the
$Z$ mass and width be defined in terms of $\bs$, the complex-valued
position of the $Z$ pole [8,9,6].
Specifically, $\bs$  is the solution of
\be
\bs = M_0^2 + \Pi^{Z Z}_T(\bs) + [\Pi^{\gamma Z}_T(\bs)]^2
/[\bs- \Pi^{\gamma \gamma}(\bs)]~,
\label{pole}
\ee
where the $\Pi$'s stand for the conventional \sesp~  If
one instead employs the PT \sesc~ the corresponding pole position
is given by
\be
\hs = M_0^2 + \hPi^{Z Z}_T(\hs) + [\hPi^{\gamma Z}(\hs)]^2
/[\hs- \hPi^{\gamma \gamma}(\hs)]~.
\label{PTpole}
\ee
As Eq.(\ref{pole}) is gauge invariant, for simplicity we evaluate it in the
$\xi_i=1$ gauge.
Subtracting Eq.(\ref{pole}) from Eq.(\ref{PTpole}),  we get
\be
\hs-\bs = \hPi^{Z Z}_T(\hs) - \Pi^{Z Z}_T(\bs) +
\left([\hPi^{\gamma Z}(\bs)]^2-[\Pi^{\gamma Z}_T(\bs)]^2\right)
/\bs + ... ~,
\label{hsminusbs}
\ee
where henceforth it is understood that the conventional \ses are evaluated
in the  $\xi_i=1$ gauge and
the ellipsis represent terms of ${\cal O}(g^6)$ and  higher.
{}From  Eq.(\ref{PTZZ}) it is easy to see that $\hs-\bs$ is of order
${\cal O}(g^4)$ or higher. Therefore, in  Eq.(\ref{hsminusbs}) we can
replace $\hPi^{Z Z}_T(\hs) \rightarrow \hPi^{Z Z}_T(\bs)$ with an error of
${\cal O}(g^6)$.
Decomposing the difference $\hPi^{Z Z}_T(\bs) - \Pi^{Z Z}_T(\bs)$
into one and two-loop parts, we have
\be
\left[\hPi^{Z Z}_T(\bs) - \Pi^{Z Z}_T(\bs) \right]_{(1)} =
-(\bs-M_0^2)4g^2c_W^2I_{WW}(\bs) =
- 4g^2c_W^2\Pi^{Z Z}_T(\bs)I_{WW}(\bs) ~,
\label{1ZZ}
\ee
where we have used Eq.(\ref{PTZZ}) and
Eq.(\ref{pole}). Furthermore
\be
\left[\hPi^{Z Z}_T(\bs) - \Pi^{Z Z}_T(\bs) \right]_{(2)} =
\left(\Pi^{Z Z}_T(\bs)\right)^P_{(2)} ~,
\label{2ZZ}
\ee
where $\left(\Pi^{Z Z}_T(\bs)\right)^P_{(2)}$ is the two-loop pinch part
to be added to the conventional \se in order to convert it into its
PT counterpart.
Extending the P-P prescription to neutral currents,
this amplitude is of the form
\be
\left(\Pi^{Z Z}_T(\bs)\right)^P_{(2)} =
C_1(q^2-M_0^2)V_2^P + C_2(q^2-M_0^2)^2B_2^P - R_{ZZ}^{(2)} ~,
\label{2loopPinch}
\ee
where $V_2^P$ and $B_2^P$ are two-loop pinch parts from vertex and box
diagrams ($C_1$ and $C_2$ are just constants) and $R_{ZZ}^{(2)}$ is
given in Eq.(\ref{R2ZZ}). Setting $q^2=\bs$, it is clear that the
first two terms in Eq.(\ref{2loopPinch}) contribute to ${\cal O}(g^6)$.
Thus we have
\bea
\hPi^{Z Z}_T(\bs) - \Pi^{Z Z}_T(\bs) = &
-4g^2c_W^2\Pi^{Z Z}_T(\bs)I_{WW} - R_{ZZ}^{(2)} \hspace*{2cm} \nonumber \\
=& 4g^2c_Ws_W\Pi^{\gamma Z}_T(\bs)I_{WW} -  4g^4c^2_Ws^2_W I^2_{WW} \bs
+ ...
\label{3ZZ}
\eea
On the other hand,
\bea
[\hPi^{\gamma Z}(\bs)]^2-[\Pi^{\gamma Z}_T(\bs)]^2 = &
 [\Pi^{\gamma Z}_T(\bs)-2g^2c_Ws_W(2\bs-M_0^2)I_{WW}]^2 -
[\Pi^{\gamma Z}_T(\bs)]^2 \nonumber \\
= & -4g^2c_Ws_W\Pi^{\gamma Z}_T(\bs)I_{WW}\bs + 4g^4c^2_Ws^2_W I^2_{WW} \bs^2
+ ...
\label{1gZ}
\eea
Combining Eqs.(\ref{hsminusbs}), (\ref{3ZZ}),  and (\ref{1gZ})
we find that the contributions of
Eqs.(\ref{3ZZ}) and (\ref{1gZ}) cancel ! Therefore,
\be
\hs-\bs =  {\cal O}(g^6)
\label{finhsminusbs}
\ee
Thus,  in analogy with the P-P results, we find that through ${\cal O}(g^4)$
the use of the PT \ses does not displace the pole position.

Defining
\be
\hA(s) = \hPi^{ZZ}_T(s) + [\hPi^{\gamma Z}(s)]^2
/[s-\hPi^{\gamma\gamma}(s)] ~,
\label{hA}
\ee
writing $\hs = \hm^2_2-i\hm_2\hGa_2$ and taking the difference between Eq.(1)
and the real part of Eq.(\ref{PTpole}), we obtain
\be
M_Z^2 = \hm^2_2- \Im m\hA'(\hm^2_2)\hm_2\hGa_2 + ...
\label{PTmass}
\ee
where $\hA'(s) = d\hA/ds$. To one-loop order, only fermionic loops
contribute to $\Im m\hA'(\hm^2_2)$. In the scaling approximation, in which
very small terms of order $m_b^2/M_Z^2$ are neglected,
\be
\Im m\hA'(\hm^2_2)=\Im m\hA(\hm^2_2)/\hm^2_2=-\hGa_2/\hm_2 ~.
\label{scaling}
\ee
Thus,
\be
M_Z^2 = \hm^2_2 + \hGa_2^2 + ...
\label{PTmass2}
\ee
Recalling Eq.(\ref{finhsminusbs}) and writing $\bs = m^2_2-im_2\Gamma_2$,
we see that, through ${\cal O}(g^4)$,
$M_Z^2 = m^2_2 + \Gamma_2^2 + ...  $. It has been previously noted that,
in terms of $m_1^2 \equiv  m^2_2 + \Gamma_2^2 $
and $\Gamma_1/m_1 \equiv \Gamma_2/m_2$,
the resonant $Z$ amplitude exhibits the $s-$dependent Breit-Wigner
resonance employed in the LEP analysis \cite{SiMass}.
Thus the PT mass , $M_Z$,
defined by  Eq.(1), can be identified with $m_1$, and therefore with
the $Z$ mass measured at LEP. We recall that a similar identification is not
consistent if, instead of $\hA$, one inserts the conventional \se $A(M_Z^2)$
in Eq.(1). In that case, for values of the gauge parameter
$\xi<1/4c^2_W$, $\Im  mA'(m^2_2)$ and, therefore $M_Z^2$, becomes gauge
dependent \cite{SiMass}.
The origin of this problem can be traced to the facts that the  conventional
\se is $\xi$-dependent and that the instability of the $Z$ forces a shift
from $M_Z^2$ to the complex valued pole position $\bs$. In fact,
it was suggested in \cite{DeSi} that these problems could in principle
be circumvented if somehow the conventional \se in the mass renormalization
condition is replaced by a  $\xi$-\ind amplitude, in such a manner that the
pole position $\bs$ is not shifted.
We have shown that  this is precisely what the
PT does.

Inserting $M_0^2 = M_Z^2 - \Re e\hA(M_Z^2)$ in the transverse $Z$ propagator
$-i/(s-M_0^2-\hA(s))$ leads to $-i/(s-M_Z^2-\hA(s)+\Re e\hA(M_Z^2))$, the
conventional mass renormalization for the  PT \se of the $Z$.
In the resonance
region,  where $s-M_Z^2 = {\cal O}(g^2M_Z^2)$, using the scaling approximation
for $\Im m\hA(s)$, the propagator can be expressed as
$$-i/[1-\Re e\hA'(M_Z^2)][s-M_Z^2+is\Gamma_Z/M_Z]~,$$ which exhibits the
characteristic $s-$dependent Breit-Wigner resonance employed in the LEP
analysis. Here $\Gamma_Z$ is the $Z$ width evaluated through ${\cal O}(g^4)$.
This expression demonstrates once more that $M_Z$, defined via Eq.(1),
can be identified with the mass measured at LEP. The factor
$[1-\Re e\hA'(M_Z^2)]^{-1}$ can be perturbatively expanded in the
amplitude's numerator where it can be combined with other radiative
corrections. Off the resonance region, where $s-M_Z^2 = {\cal O}(M_Z^2)$,
one can expand the propagator in powers of
$[\hA(s)-\Re e\hA(M_Z^2)]/(s-M_Z^2)$. Alternatively, one can retain
$\hA(s)-\Re e\hA(M_Z^2)$ in the propagator's denominator, in which case one
needs a renormalization prescription to eliminate the remaining UV divergences
in $\hA(s)-\Re e\hA(M_Z^2)$. In the $\bar{MS}$-scheme, for instance, it is
natural to retain $(\hA(s)-\Re e\hA(M_Z^2))_{\bar{MS}}$, where the
$\bar{MS}$ subscript indicates that the $\bar{MS}$ renormalization has
been carried out.

We emphasize that the PT \ses are $\xi$-\ind, treat the fermionic and
bosonic contributions on an equal footing and, as explained before, do not
displace the complex-valued position of the pole, as we showed explicitly
up to ${\cal O}(g^4)$. The ultraviolet divergences reside in the one-loop
\ses and can be absorbed in the renormalization of the bare couplings.
For $|q^2| >> M_Z^2$ the PT \ses satisfy the RGE. Thus for large $|q^2|$,
they can be interpreted as factors that transform the bare into running
couplings.
In summary, the PT approach provides a convenient and
rather elegant framework to discuss important neutral current amplitudes.

\section{Aknowledgements}
The work of K.P. was supported in part by the European Union grant
CHRX-CT 93-0319, a Sokol Travel/Research Award and a Dean's Summer Fellowship
for Preliminary PhD. Research. The research of A.S. was supported in part
by the National Science Foundation under grant No. PHY-9313781.

\section{Figure Captions}
{\bf Figure 1} ~: One -loop \se diagrams for neutral currents
amplitudes in the SM ($R_{\xi}$ gauges ). The \ses are the conventional
$R_{\xi}$ amplitudes and the unoriented solid lines stand for the
$R_{\xi}$ bosonic propagators.

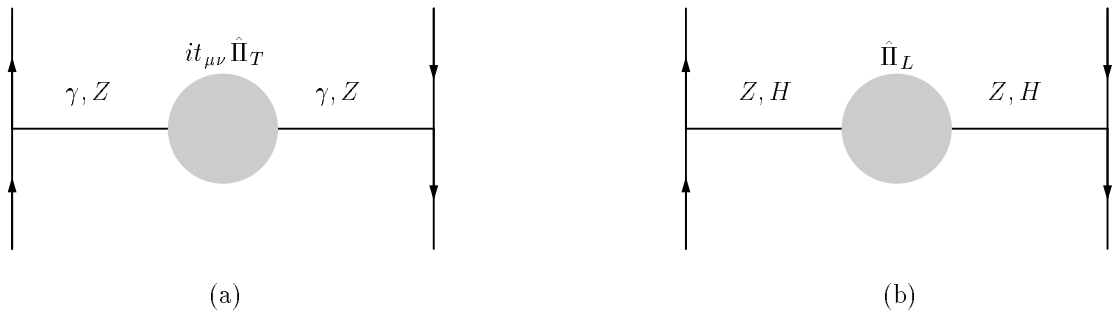
{\bf Figure 2} ~:  One-loop \se diagrams in the PT approach. The four
diagrams in  (a) correspond to the transverse PT self-energies of the
vector bosons (proportional to $t_{\mu\nu}$). The four diagrams in (b)
involve $\ell_{\mu\nu}\hPi^{ZZ}_L,$~ $\hPi^{ZH}_{\mu},$ and $\hPi^{HH}$.

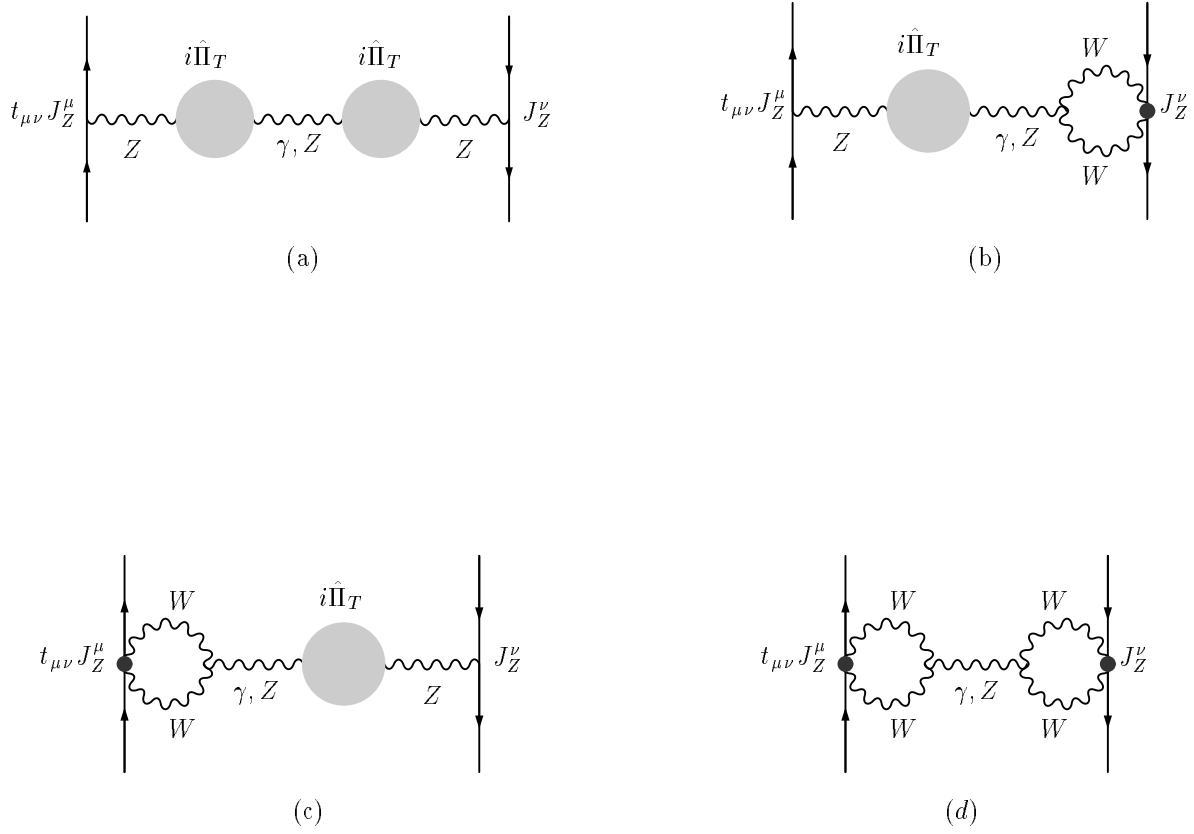
{\bf Figure 3} ~: Two-loop chains of transverse PT \ses and
a class of related
pinch parts in the PT approach. Only contributions proportional to
the external currents $J^{\mu}_Z$ and  $J^{\nu}_Z$ are shown.

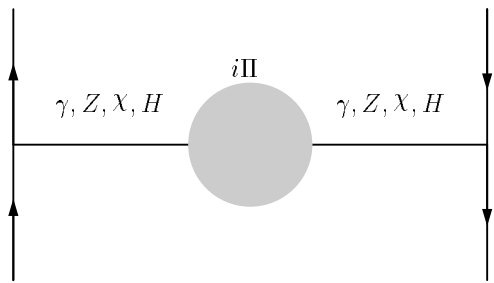
\end{document}